\documentclass[useAMS,usenatbib,12pt]{mn2e}
\usepackage[dvipdf]{graphicx,color}
\usepackage[latin1]{inputenc}
\usepackage{graphics}
\usepackage{amsfonts}
\usepackage{amsmath}
\usepackage{multicol}
\usepackage{layout}
\usepackage{amssymb}
\usepackage{epsfig}
\usepackage{hyperref}
\DeclareGraphicsExtensions{.eps}

\title[Mapping Star Formation in Galaxy Interactions]{Mapping galaxy encounters in numerical simulations: \\The spatial extent of induced star formation}

\author[J. Moreno et al.]{Jorge Moreno$^{1,2,3,\dagger}$,  Paul Torrey$^{2,4}$, Sara L. Ellison$^{3}$, David R. Patton$^{5}$, Asa F. L. Bluck$^{3}$, \newauthor Gunjan Bansal$^{3,6,7}$ \& Lars Hernquist$^{8}$ \\
$^{1}$Department of Physics and Astronomy, California State Polytechnic University, Pomona, Pomona, CA 91768, USA\\
$^{2}$TAPIR, Mailcode 350-17, California Institute of Technology, Pasadena, CA 91125, USA\\
$^{3}$Department of Physics and Astronomy, University of Victoria, Finnerty Road, Victoria, British Columbia, V8P 1A1, Canada\\
$^{4}$Department of Physics, Kavli Institute for Astrophysics and Space Research, Massachusetts Institute of Technology, Cambridge, MA 02139, USA\\
$^{5}$Department of Physics and Astronomy, Trent University, 1600 West Bank Drive, Peterborough, Ontario, K9J 7B8, Canada\\
$^{6}$Department of Physics and Astrophysics, University of Delhi, Delhi 110007, India\\
$^{7}$Leiden Observatory, Leiden University, P.O. Box 9513, 2300 RA Leiden, the Netherlands\\
$^{8}$Harvard-Smithsonian Center for Astrophysics, 60 Garden Street, Cambridge, MA, 02138, USA\\
$^{\dagger}$CITA National Fellow}

\begin{document}
\date{}
\pagerange{\pageref{firstpage}--
\pageref{lastpage}} \pubyear{2015}
\maketitle
\label{firstpage}


\begin{abstract}

We employ a suite of 75 simulations of galaxies in idealised major mergers (stellar mass ratio $\sim$2.5:1), with a wide range of orbital parameters, to investigate the spatial extent of interaction-induced star formation. Although the total star formation in galaxy encounters is generally elevated relative to isolated galaxies, we find that this elevation is a combination of intense enhancements within the central kpc and moderately suppressed activity at larger galacto-centric radii. The radial dependence of the star formation enhancement is stronger in the less massive galaxy than in the primary, and is also more pronounced in mergers of more closely aligned disc spin orientations. Conversely, these trends are almost entirely independent of the encounter's impact parameter and orbital eccentricity. Our predictions of the radial dependence of triggered star formation, and specifically the suppression of star formation beyond kpc-scales, will be testable with the next generation of integral-field spectroscopic surveys.

\end{abstract}



\section{Introduction}\label{secIntro}

Galaxy encounters are recognized as a leading mechanism for triggering star formation \citep{sanders96}. Numerical simulations support this picture: interaction-driven non-axisymmetric gravitational torques funnel copious amounts of gas into the central regions, fueling powerful bursts of star formation \citep{hernquist89,barnes91,barnes96}. However, this does not necessarily mean that starbursts are ubiquitous in interacting galaxies. Triggering depends on many factors, including the specific merging geometry \citep{cox06,cox08,dimatteo07,dimatteo08,torrey12} and the properties of the progenitor galaxies \citep{mihos96,springel00,springel05b}.  Even if triggering occurs, the observability time-scale of the burst may be shorter than the duration of the interaction. In other words, a galaxy in an encounter may be identified as normal (or even passive) if observed long after the peak of activity.

Despite these issues, surveys detect (on average) elevated levels of star formation in galaxies with close companions \citep{barton00,barton03,lambas03,alonso06,woods06,ellison08,patton11,scudder12,scott14}. Relevant related works include investigations on the role of environment \citep{alonso04,ellison10,alonso12,kampczyk13,ellison13} --  and extensions to wider separations \citep[out to $\sim$150 kpc;][]{patton13} and high redshift \citep{lin07,freedman10,hwang11,wong11,kampczyk13}.

To shed light on how merger-induced star formation unfolds, it is interesting to ask where the transformation of gas into stars is most efficient. Several works address this question by focusing on single individual systems in detail (e.g., the Antennae -- see Section~\ref{secDiscussion}). Unfortunately, large surveys are not yet able to provide a statistical view of the spatial distribution of star formation in galaxies with close companions. A crude attempt is performed by \cite{ellison13}, who measure the star formation rate (SFR) in apertures (`fibres') centred on galaxies, and compare it to the total SFR of those galaxies \citep[see also][]{patton11,scudder12,patton13}. They show that fibre SFR is more elevated (relative to control galaxies without close companions) than total SFR - suggesting that centrally-concentrated star formation in interactions is common. Unfortunately, using a fixed angular fibre means that different portions of the target galaxies are covered -- i.e., coverage depends on how large galaxies appear on the sky (i.e., their distance to us). This obstacle severely limits our ability to properly quantify the spatial concentration of star formation in interacting galaxies.

Ideally, surveys equipped with the ability to map the location and kinematics of star forming regions would provide better clues on how this process unfolds. \cite{kewley10}, \cite{rupke10} and \cite{rosa14} use HII-regions to find that interacting galaxies have shallow metallicity gradients, which is consistent with inflow of gas that ultimately ignites starbursts. Similarly, \cite{rich12} use integral field spectroscopy (IFS) to create detailed metallicity maps in interacting systems. More direct approaches include \cite{knapen09}, who use H$\alpha$ imaging to estimate SFR profiles in mergers -- and \cite{bellocchi13}, who employ IFS to determine changes in the 2D kinematic behaviour of ionised gas (H$\alpha$) as the merging sequence advances.  Their main limitation, however, is that the samples considered often contain too few galaxies. In other words, it is not clear that the information we infer from just a handful of cases is universal.

\cite{schmidt13} are the first to attempt using a relatively large sample: 60 mergers at $z\sim1.5$, visually-selected with \textsc{3d-hst}. These authors use near-infrared slitless spectroscopy to measure the spatial extent of star formation (via H$\alpha$ and [OIII] emission-line maps). Unfortunately, their scheme is very crude:  it only checks if star formation occurs in a single galaxy, in both, or in the region connecting the two galaxies. In this paper, we underscore the need for observations capable of measuring resolved star-formation maps in interacting galaxies.

In this direction, the emergence of large multi-Integral-Field-Unit (IFU) spectroscopic programmes holds considerable promise. Surveys like \textsc{califa}\footnote{The Calar Alto Legacy Integral Field spectroscopy Area survey (\url{http://califa.caha.es})} \citep{sanchez14}, \textsc{sami}\footnote{The Sydney-Australia-Astronomical-Observatory Muti-object Integral-Field Spectrograph (\url{http://sami-survey.org})} \citep{croom12}, \textsc{m}{\small a}\textsc{nga}\footnote{Mapping Nearby Galaxies at the Apache Point Observatory (\url{https://www.sdss3.org/future/manga.php})} \citep{bundy15}, and the future \textsc{hector} survey \citep{lawrence12} will soon be able to analyse large samples of interacting galaxies with exquisite spatial detail. Indeed, both \textsc{califa} (Barrera-Ballesteros et al., in prep) and \textsc{sami} are already on their way to analysing their respective samples of interacting galaxies (Jorge Barreda-Ballesteros and Iraklis Konstantopoulos, private communications).  Beyond spectroscopic mapping, mid-infrared imaging surveys like \textsc{s4g}\footnote{Spitzer Survey of Stellar Structure in Galaxies (\url{http://www.cv.nrao.edu/~ksheth/S4G})} \citep{sheth10} will also provide clues on the merging sequence, and its impact on the structure of galaxies \citep{knapen14}.

With these surveys in sight, the time is right to conduct resolved spatial studies of star formation in interacting galaxies with numerical simulations. The aim of this paper is to answer the following question: is star formation in interacting galaxies nuclear or extended? Also, which orbital parameters govern the spatial extent of star formation?

This paper is organised as follows. We present our methods in Section~\ref{secMethods}. Section~\ref{secFiducial} describes a case study, and Section~\ref{secOrbits} generalises to other merger configurations. We discuss our findings in Section~\ref{secDiscussion}, and summarise in Section~\ref{secConclusions}.


\section{Methods \& Definitions}\label{secMethods}

\subsection{The Model}\label{subsecModel}

We use the smoothed-particle hydrodynamics (SPH) code \textsc{gadget-3}~\citep{springel05gadget} to run idealised galaxy merger simulations. These include gravity, hydrodynamics, radiative gas cooling~\citep{katz96}, star formation with associated feedback~\citep{springel03}, and supermassive black hole growth and feedback~\citep{dimatteo05}. {This model employs an equation of state parameter of $q=0.3$ to handle the pressurization of dense, star forming gas}.  Regarding our formulation of SPH~\citep{springel02}, we do not expect our results to be sensitive to details of the hydro solver~\citep{hayward14}.  

{This model has the advantage of being well numerically converged.  By suppressing gas fragmentation, it ensures that star formation proceeds at an efficiency consistent with the \cite{kennicutt98} (KS) relation.  Therefore, the results presented in this paper are mostly dependent on our ability to resolve the mechanisms that redistribute the gas throughout the galaxy (e.g., bars, arms, tidal forces, gas shock heating, etc.) with the local SFRs then being determined by an enforced volumetric KS relation.  

Other models in the literature \citep[e.g.,][]{teyssier10,powell13,renaud14,renaud15} attempt to determine the SFR efficiency in mergers by resolving parsec scale physical processes.  Here, we only hope to understand where the gas has moved during the merger, and therefore where we would expect SFR to occur. It is therefore worth cautioning that if our adopted KS relation is invalid in merging systems, then this could impact our results.  However, a significant violation of the KS relation would be required to change our qualitative conclusions.}

\subsection{Galaxy Merger Simulations}\label{subsecGalMer}

{In this paper, we employ the simulation suite first discussed in \cite{patton13}. This merger suite differs from that presented in \cite{torrey12} only in the adopted initial conditions and orbital parameters.  The same simulation code and physics modules are adopted in both.  As such, we direct the reader to \cite{torrey12} for a more complete description of our employed physics modules.}

We focus exclusively on mergers with stellar mass ratios $\sim$2.5:1 -- where, initially, the larger ({\bf primary}) galaxy has stellar mass of $M_*=1.4 \times 10^{10} M_\odot$ and the smaller ({\bf secondary}) galaxy has a stellar mass of $M_* = 5.7\times10^9 M_\odot$.  These specific choices represent typical values in the Sloan Digital Sky Survey galaxy-pair catalogue of \cite{patton13} -- and are also commonplace in the cosmological galaxy-pair catalogue of \cite{moreno13}, drawn from the Millennium Simulation \citep{springel05ms}. 

For both galaxies, we adopt an initial stellar bulge-to-disc ratio of $M_{{\rm bulge}} / M_{{\rm disc}} = 0.24$, as in \cite{patton13}. The initial galaxies are set up following the analytic work of~\citet{mo98}, via the procedure outlined in~\cite{springel05}. Simulated bulges follow a~\citet{hernquist90} profile -- and simulated discs have scale lengths of 2.1 kpc (for the more massive galaxy) and 1.5 kpc (for the less massive galaxy), for both the gaseous and stellar components. The initial gas fraction is set at $f_{\rm gas} = M_{{\rm gas}} / M_{{\rm disc}} = 0.25$, consistent with observations \citep{catinella12}. 

{Each run has $\sim$2.5$\times10^6$ baryon particles, yielding a baryon mass resolution of $M_{\rm b}\sim10^4$, with a Plummer equivalent gravitational softening length of 50 pc.  Each simulation has 2$\times10^7$ dark matter particles, yielding a dark matter mass resolution of $M_{\rm DM} = 2.5\times10^6$ M$_{\odot}$, with a Plummer equivalent gravitational softening length of 200.}

Our merger simulations are initiated by placing two otherwise stable galaxies on an interacting orbit. We consider 75 merger simulations in total in this paper, consisting of 25 different orbital configurations (variations in the orbital energy and angular momentum) with three different alignments of the galaxy's angular momentum relative to the plane of the merger (see below). The 25 orbital configurations are built by considering orbital eccentricities $\epsilon=\{0.85, 0.90, 0.95, 1.0, 1.05 \}$ and Keplerian-inferred impact parameters of $b=\{2, 4, 8, 12, 16\}$ kpc. These choices are consistent with cosmological simulations \citep{khochfar06}.

\begin{figure}
 \centering
 \includegraphics[width=\hsize]{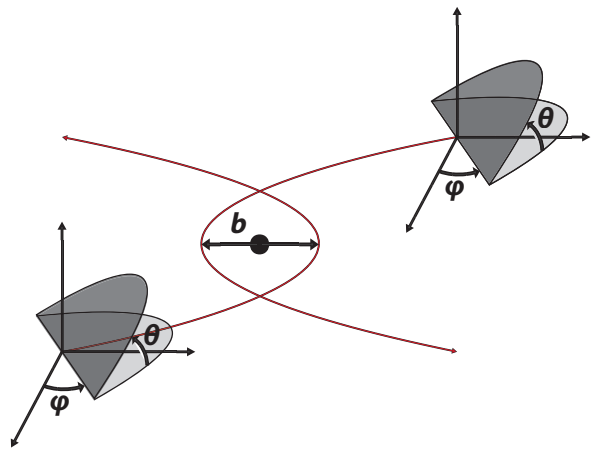}
 \caption{Schematic representation of angles defining the relative orientation of our merging galaxies. See Table~\ref{tableAngles} for the specific angles employed in this paper. Figure adapted from \citet{torrey12} -- see their Figure~6. \copyright AAS. Reproduced with permission.}
 \label{figDiskOrientation}
\end{figure}

We employ the following three merger orientations: these are the ``e", ``f", and ``k" orientations drawn from~\citet{robertson06}, and summarised in Table~\ref{tableAngles}. These orientations are selected to represent two strongly aligned discs (``e"-orientation), two nearly perpendicular discs (``f"-orientation), and two nearly anti-aligned discs (``k"-orientation). See Figure~\ref{figDiskOrientation} for a schematic description \citep[adapted from][]{torrey12}.  

\begin{center}
\begin{table}
\begin{tabular}{|| c |l c || c || c || c || c ||}
\hline
Orientation & $\,\,\,\,\,\,\,\,\,\,\phi_{1}$ & $\theta_{1}$ & $\phi_{2}$ & $\theta_{2}$ \\
Identifier & [degrees] & [degrees] & [degrees] & [degrees] \\

\hline \hline 
{\rm e} & $\,\,\,\,\,\,\,\,\,$60 & 30 & 45 & -30 \\
{\rm f} & $\,\,\,\,\,\,\,\,\,$60 & 60 & 0 & 150  \\
{\rm k} & $\,\,\,\,\,\,\,$-30 & -109 & -30 & 71 \\
\hline 
\end{tabular}
\caption{The merger orientations considered in this paper, drawn from \citet{robertson06}, and described in Figure~\ref{figDiskOrientation}. 
}
\label{tableAngles}
\end{table}
\end{center}

\begin{figure*}
 \centering
 \includegraphics[width=\hsize]{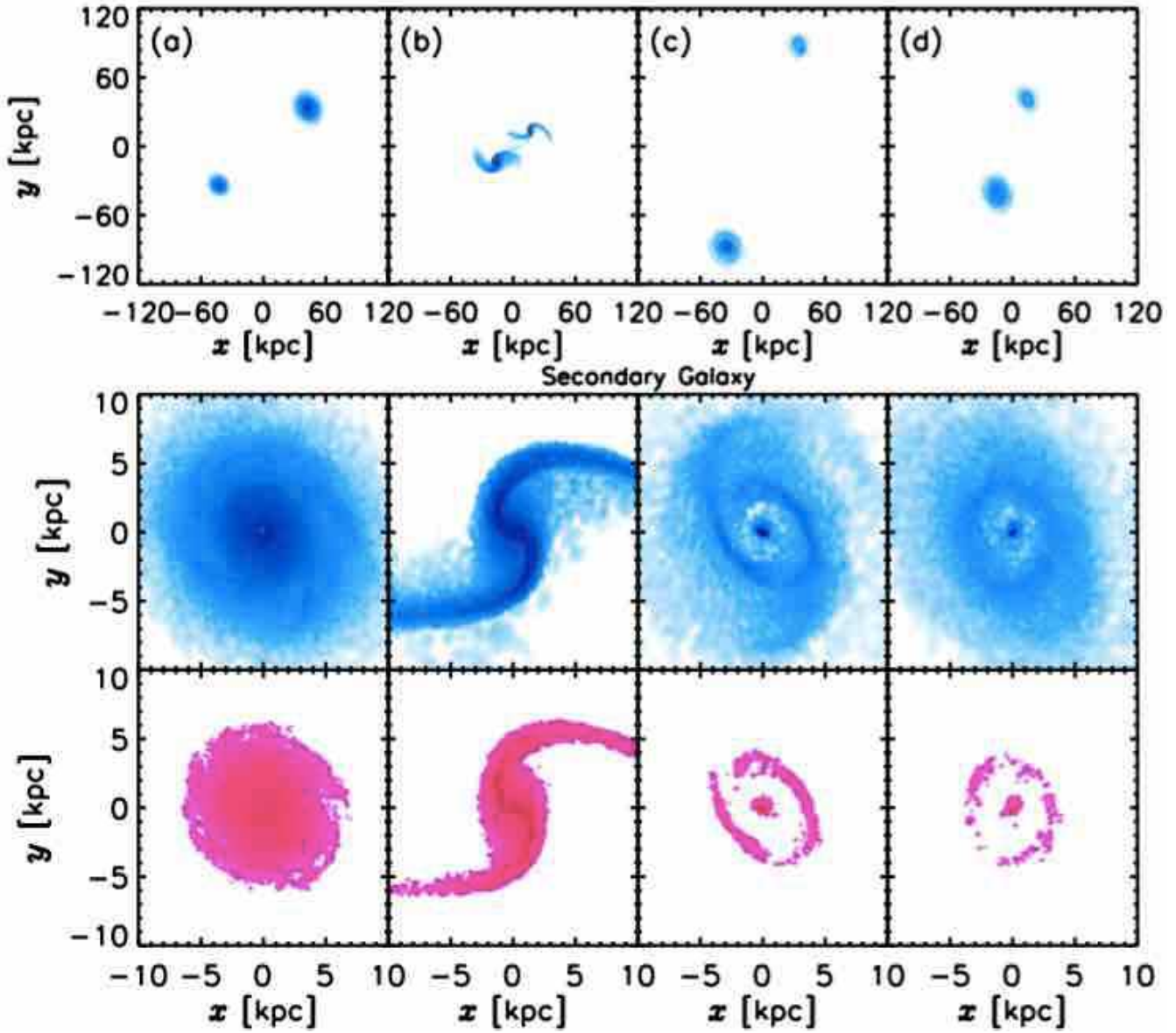}
 \caption{Density maps: gas ({\it top} and {\it middle}, blue) and star-forming gas ({\it bottom}, pink). Upper row displays 120 kpc $\times$ 120 kpc stamps of the interacting system. Middle and lower rows are centred on the secondary (smaller) galaxy, displayed on 20 kpc $\times$ 20 kpc stamps. Columns (left-to-right) show the following stages: incoming (a); first passage (b); apocentre (c); and second approach (d). First passage produces tidal tails, followed by a nuclear starburst and a mild star-forming contribution at larger galacto-centric radii.}
 \label{figMaps}
\end{figure*}

\subsection{Galaxy Membership \& the Interacting Phase}\label{subsecMembership}

The central goal of this paper is to map star formation in interacting galaxies, and compare to equivalent maps of their isolated counterparts. A nuisance of this procedure is how to assign SPH particles to each of the two galaxies. Some works assign membership by seeking the nearest supermassive black hole, which is treated as a proxy for the centre of its host galaxy \citep[][]{torrey12,patton13}. This procedure becomes particularly tricky when the separation between the two galaxies is smaller than their typical sizes. 

Our approach is slightly different. For each galaxy, we only focus on the SPH particles contained within a sphere of 10-kpc radius centred around the corresponding supermassive black hole. We adopt this radius because our two galaxies in isolation form their stars entirely within 10-kpc. We also checked that the fraction of star formation taking place outside such spheres in the interacting case is negligible for all of our runs. This approach is powerful because it allows us to compare galaxies with companions to their isolated equivalents directly, on a region-by-region basis. Breakdown occurs when the two spheres overlap: SPH particles are assigned to two galaxies simultaneously, leading to double counting. To avoid this complication, we ignore those few snapshots where the separation between the black holes is less than 20 kpc. 

In this paper, we are primarily interested in the stages of merging where the two distinct galaxies can be identified unequivocally. For this reason, we focus exclusively on the {\bf interacting phase}: the period between first and second pericentric passage (with the provision that separation is greater than 20 kpc). This adopted approach facilitates comparing the spatial extent of star formation in interacting galaxies to that in their isolated counterparts.


\section{Results: A Case Study}\label{secFiducial}

This section describes a {\bf case study} in full detail (eccentricity $\epsilon=1.05$, impact parameter $b=16$ kpc). This choice is not meant to be average. Instead, our goal is to maximise the duration of the interacting phase (prior to coalescence).  All other runs exhibit qualitatively similar features, except that they are always interrupted by merging at an earlier time. We first focus primarily on the ``e'' orientation. See Section~\ref{subsecOrientation} for other orientations.

\subsection{Mapping Star Formation}\label{subsecFidMap}

Figure~\ref{figMaps} shows density maps of the gas (blue) and star forming gas (pink) for our case study. The upper row shows 120 kpc $\times$ 120 kpc stamps, depicting the evolution of the interaction on extra galactic scales. The middle and bottom rows show 20 kpc $\times$ 20 kpc stamps centred on the secondary (smaller) galaxy. Star-forming gas (bottom row) only traces the densest gas (middle row, in blue), as expected in our \cite{kennicutt98} based model \citep{springel03}.

Columns (a)-(d) represent various stages of interaction:

\begin{itemize}

\item{\bf (a) Incoming Phase:} Before the interaction. Left alone, these galaxies retain their original morphology, and exhibit declining star formation as a function of time.

\item{\bf (b) First Pericentric Passage:} The galaxies exhibit short-lived tidal tails.

\item{\bf (c) Apocentre:} Gas density is increased in the centre and suppressed in the outskirts. This produces a strong nuclear burst and mild off-nuclear star formation.

\item{\bf (d) Second Approach:} Morphology is very similar to that at apocentre (c), but with lower levels of star formation.

\end{itemize}

Only columns (b)-(d) correspond to the interacting phase (Section~\ref{subsecMembership}). Merging and post-coalescence phases are omitted. The incoming phase (column a) is included for comparison (identical to the isolated case).  It is evident that the spatial distribution of star formation in interacting galaxies has highly complex morphology. (Section~\ref{secDiscussion} briefly discusses the ring-like feature in columns c and d.)

\begin{figure}
 \centering
 \includegraphics[width=\hsize]{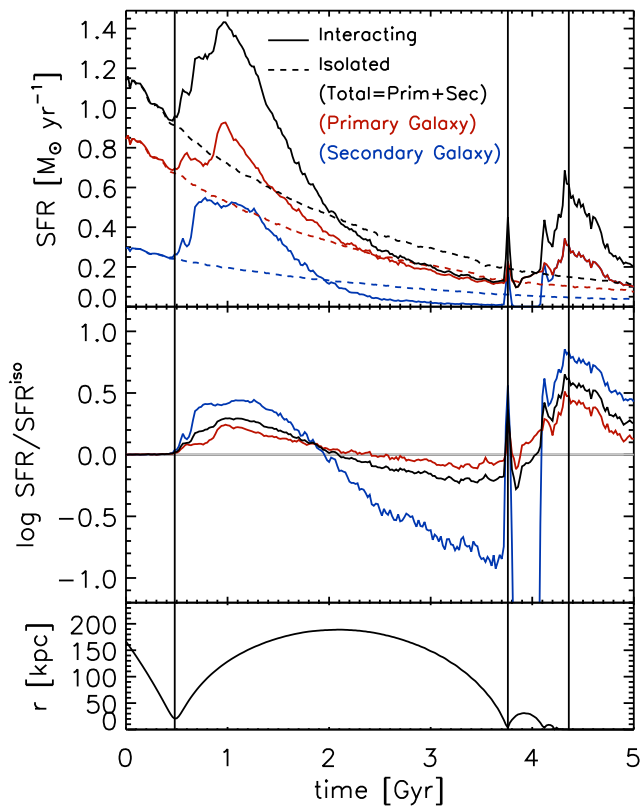}
 \caption{Global star formation rate (SFR) versus time (case study). Vertical lines delimit the stages of merging: first approach, interaction, coalescence, and post-coalescence. Red (blue) refers to the primary (secondary) galaxy, and black to the arithmetic sum of the two. Solid (dashed) curves indicate interacting (isolated) galaxies. {\it Top:} Global SFR. {\it Middle:} The logarithm of the SFR enhancement (SFR in interaction divided by SFR in isolation). Grey horizontal line indicates SFR enhancement equals unity. {\it Bottom:} Orbital separation. The galaxies experience two bursts, one in the interacting phase, the other at coalescence. The secondary galaxy exhibits weaker SFR and stronger SFR-enhancement (and suppression) than the primary.} 
 \label{figGlobalFiducial}
\end{figure}

\begin{figure}
 \centering
 \includegraphics[width=\hsize]{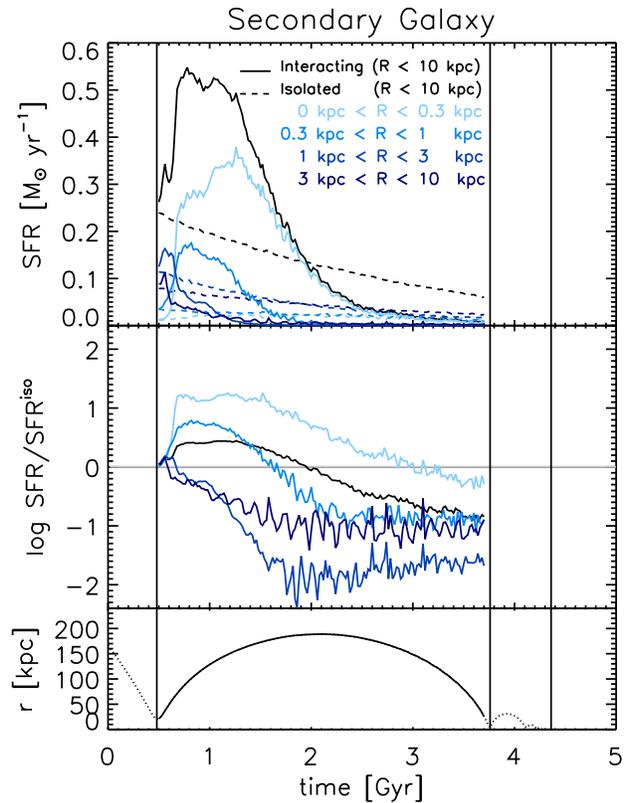}
 \caption{SFR versus time, in spherical shells of radii $[0-0.3]$, $[0.3-1]$, $[1-3]$ and $[1-10]$ kpc (light-to-dark blue), and in total (black). Secondary galaxy, case study, interacting phase only. Solid (dashed) refers to the interacting (isolated) galaxy. {\it Top:} SFR per shell. {\it Middle} The logarithm of enhanced-SFR per shell. Grey horizontal line indicates SFR enhancement equals unity. {\it Bottom:} Orbital separation (dotted refers to snapshots not considered). The innermost region dominates the total SFR, especially late in the interaction. Enhancement there is sustained, reaching factors higher than 10. The second shell is strongly enhanced for $\sim$1 Gyr, followed by suppression. The outermost shells are briefly enhanced, and quickly suppressed thereafter.}
  \label{figLocalFiducial}
\end{figure}

\subsection{The Evolution of Global Star Formation}\label{subsecFidGlobal}
 
Figure~\ref{figGlobalFiducial} shows SFR (top), SFR enhancement (middle), and separation (bottom) as a function of time (case study). {\bf SFR enhancement} is defined as the SFR of the interacting galaxy divided by the SFR of its isolated counterpart. The solid vertical lines demarcate the different stages of merging (left-to-right): incoming, interacting, coalescing, and post-coalescence. The red (blue) curves refer to the primary (secondary) galaxy, and the black curves represent their arithmetic sum (labelled `total' in the Figure). Solid (dashed) lines represent interacting (isolated) galaxies. 

Left alone, the two galaxies in isolation experience a simple decaying star formation history. In interaction, on the other hand, the galaxies experience two bursts of star formation: one between first and second pericentric passage, and another when the final merger occurs. 

The Figure illustrates the nuisances described in Section~\ref{subsecMembership}. The spikes at second pericentric passage signal the overlapping of the two 10-kpc-radius spheres encompassing each galaxy. For each galaxy, the SFR boost is caused by contamination from the `invading' companion. This is particularly dramatic during coalescence, as the two spheres merge, causing the red and blue curves to converge. Our aim here is to map star formation in the interacting phase, where the two distinct galaxies are clearly identified. For the rest of this paper, these nuisances are avoided and ignored.

Our two metrics, SFR and SFR enhancement, are complementary. The larger galaxy has higher SFR and lower SFR-enhancement (compare red to blue). In other words, the smaller galaxy makes fewer stars, but is more susceptible to the encounter because of the relative tidal forces acting between the galaxies \citep[e.g.,][]{donghia10}. Notice that the interacting phase exhibits both enhancement and suppression. This is only true for configurations with sufficiently-long interacting timescales. In general, this trend is interrupted by merging.

\subsection{The Spatial Evolution of Star Formation}\label{subsecFidLocal}

Figure~\ref{figLocalFiducial} shows the time evolution of SFR (top) and SFR enhancement (middle) in concentric spherical shells centred around the supermassive black hole of the secondary galaxy. We only discuss this galaxy for the sake of brevity. We focus on the following radii: $[0-0.3]$, $[0.3-1]$, $[1-3]$, and $[3-10]$ kpc (light-to-dark blue). The total SFR (within 10 kpc) is shown in black. Solid (dashed) curves refer to the interacting (isolated) case. The bottom panel shows the orbital separation as a function of time. The dotted curve represents periods excluded from the analysis (outside the interacting phase and overlapping 10-kpc spheres, see Section~\ref{subsecMembership}).

The innermost region (with distance $R < 0.3$ kpc from the centre) exhibits the strongest levels of triggered star formation. It takes only $\sim$0.5 Gyr after first pericentric passage for this region to account for half of the star formation in the galaxy. By $\sim$1 Gyr after first passage, nearly all of the star formation is taking place in this region (compare light blue and black curves). This region is enhanced up to factors of $\sim$15 (compare to the global SFR-enhancement of $\sim$2-3, black curve). The second shell ($0.3 < R < 1$ kpc) experiences a slightly weaker burst -- with shorter duration ($\sim$1 Gyr), followed by suppression. SFR in this region is enhanced by factors of $\sim$5-6. The two outermost shells experience a brief (and weak) episode of SFR enhancement, and are quickly suppressed. In particular, the shell next to last ($1  < R < 3$ kpc) experiences the strongest suppression -- two orders of magnitude below the isolated case.

In summary, the central regions experience stronger and longer periods of SFR enhancement, whilst activity in the outskirts is largely suppressed during the encounter.  The increase in central star formation is due to the redistribution of gas is caused by non-axisymmetric tidal torques produced by the interaction, which leads to high gas density concentrated in the centre (Figure~\ref{figMaps}), and which has been previously seen in simulations \citep[e.g.,][]{mihos96,iono04}. This increase in nuclear SFR is in line with recent simulations by \cite{hopkins13} and \cite{renaud15} -- although those analyses do not compare against SFR in isolated galaxies. To our knowledge, we are the first to report {\it suppression} of star formation at galacto-centric radii.  In Moreno et al. (in prep) we investigate the mechanisms responsible for this suppression, and leave a discussion of that effect for this forth-coming paper.


\begin{figure}
 \centering
 \includegraphics[width=\hsize]{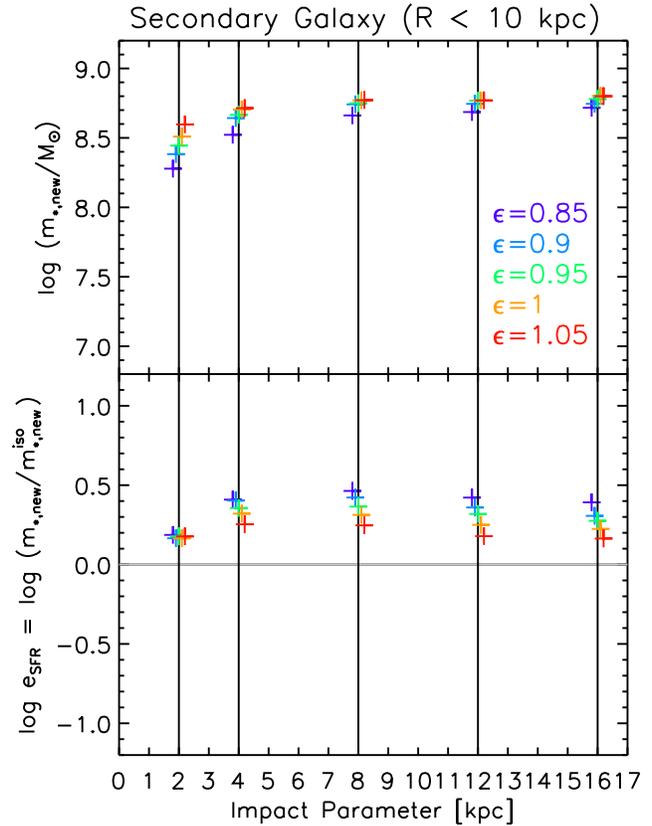}
 \caption{Global stellar mass, $m_{*,\,{\rm new}}$ (equation~\ref{eqnmnew}, {\it top}) and SFR efficiency, ${\rm e}_{\rm SFR}$ (equation~\ref{eqneff}, {\it bottom}), for secondary galaxy in the interacting phase. Colours indicate eccentricities (purple-to-red): $\epsilon=\{0.85, 0.90, 0.95, 1.0, 1.05 \}$. Vertical lines refer to these impact parameters: $b=\{2, 4, 8, 12, 16\}$ kpc. Symbols are slightly offset for clarity. For all 25 orbits, more stars are made in the interacting galaxy than in its isolated counterpart (all symbols in the bottom panel are above the grey horizontal line).}  
 \label{figGlobalEIP}
\end{figure}

\section{Results: 75 Merger Simulations}\label{secOrbits}

\begin{figure*}
 \centering
   \includegraphics[width=\columnwidth]{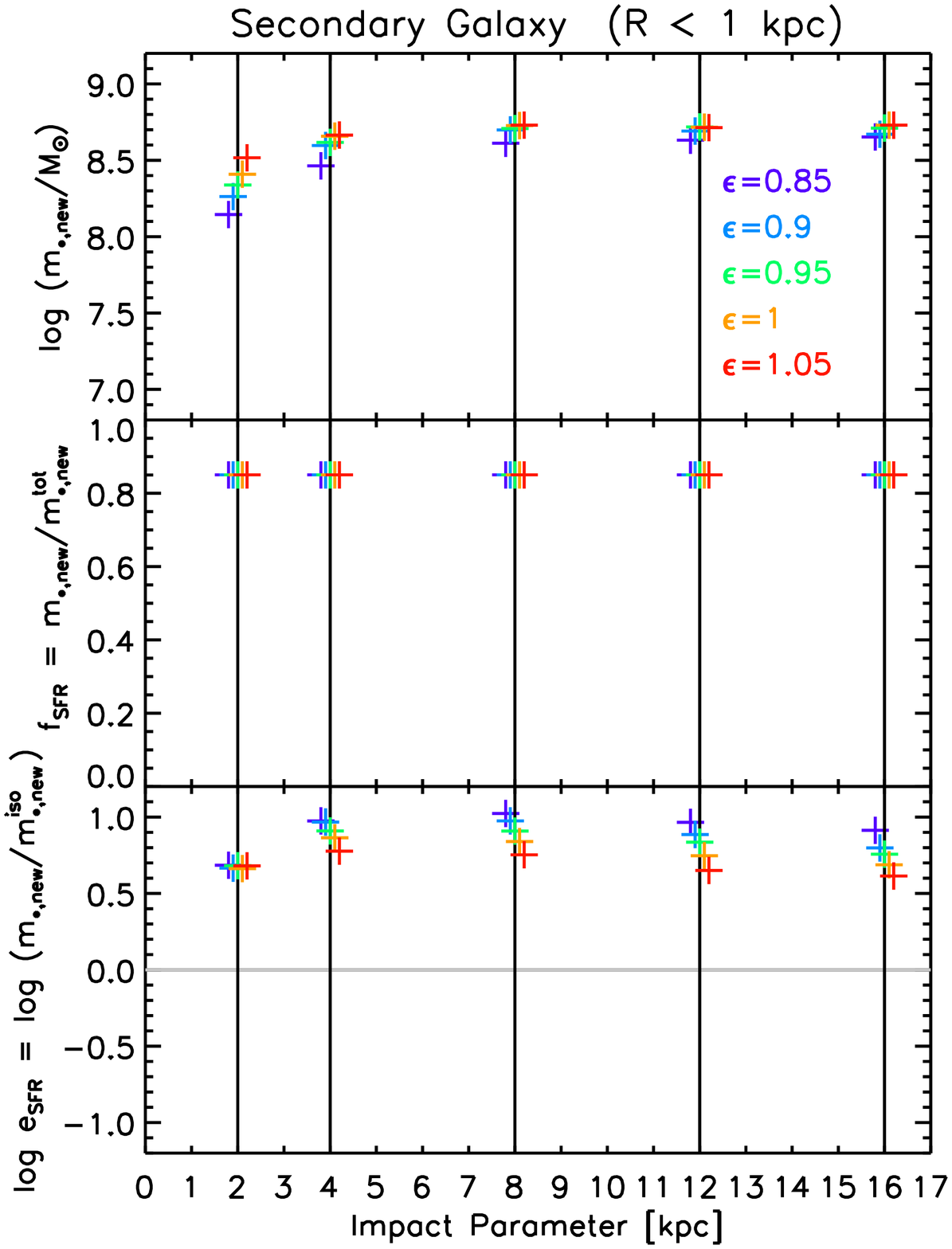}
 \includegraphics[width=\columnwidth]{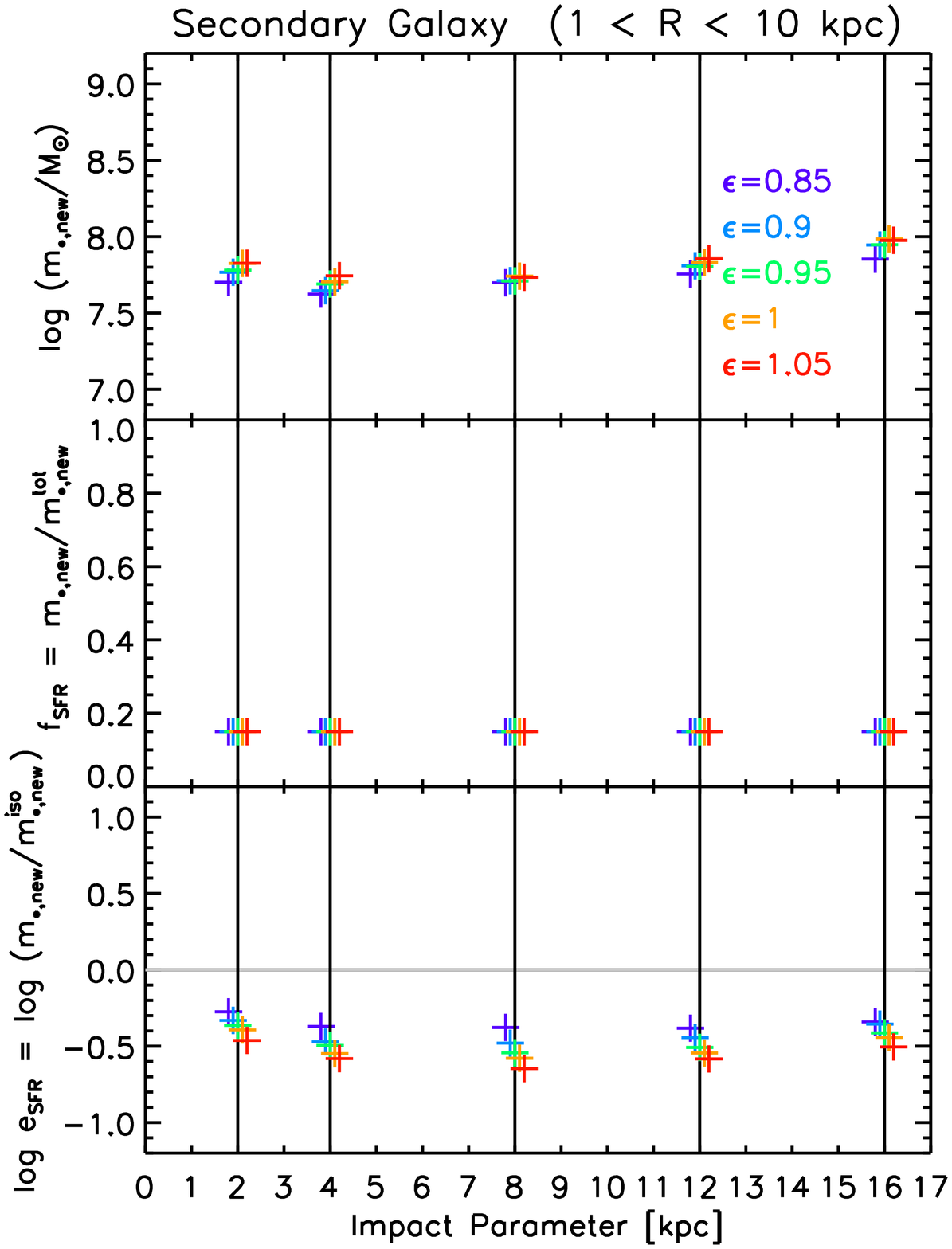}
   \caption{Stellar mass ({\it top}), SFR fraction (equation~\ref{eqnfrac}, {\it middle}) and SFR efficiency ({\it bottom}) in the secondary galaxy: nuclear ($R < 1$ kpc) versus off-nuclear ($1 < R < 10$ kpc) contributions ({\it left} versus {\it right}). Colours, symbols, and vertical lines as in Figure~\ref{figGlobalEIP}. Grey horizontal line refers to SFR efficiency of unity. Star formation is split into $\sim$87 \% (nucleus) and $\sim$13 \% (outskirts). SFR efficiency is enhanced (suppressed) inside (outside) the central kpc.}
 \label{figLocalEIP}
\end{figure*}

\subsection{Eccentricity \& Impact Parameter}\label{subsecEIP}

We now explore various eccentricities and impact parameters: $\epsilon=\{0.85, 0.90, 0.95, 1.0, 1.05 \}$, $b=\{2, 4, 8, 12, 16\}$ kpc.

\subsubsection{Global Star Formation Efficiency}\label{subsubsecGlobalEff}

It is impractical to describe every possible merger in our suite with the same level of detail devoted to our case study (Section~\ref{secFiducial}). Instead, we adopt the {\bf integrated star formation}, which is equivalent to the total mass in stars created during the interacting phase (Section~\ref{subsecMembership}):
\begin{equation}
m_{*,\,{\rm new}}=\int_{\rm interacting \, phase} SFR(t) \,{\rm d}\,t,
\label{eqnmnew}
\end{equation}
This quantity is equivalent to the `ISFR' of \cite{dimatteo07}, except that we integrate exclusively over the interacting period. This is also equivalent to the mass in gas that turns into stars \citep{cox08}. Alternatively, we could adopt ${\rm SFR}_{\rm max}$, the maximum SFR. However, the intermittent nature of star formation renders this quantity inadequate because it depends strongly on our time-step choice.

We also introduce the {\bf star formation efficiency}:
\begin{equation}
{\rm e}_{\rm SFR}=\frac{m_{*,\,{\rm new}}}{m^{\rm iso}_{*,\,{\rm new}}},
\label{eqneff}
\end{equation}
where $m_{*,\,{\rm new}}$ is defined in equation~(\ref{eqnmnew}), and $m^{\rm iso}_{*,\,{\rm new}}$ is its analogue in isolation. This quantity is similar to the `burst efficiency' of \cite{cox08} -- except that (1) we only consider the interacting phase; (2) they compute a difference where we compute a ratio; and (3) they use the total SFR of the two galaxies, whilst we use it for individual galaxies (this section), and for subregions therein (Section~\ref{subsubsecLocalEff} below).

Figure~\ref{figGlobalEIP} shows the integrated star formation (top) and star formation efficiency (bottom). The solid vertical lines mark impact parameters (the symbols are offset for clarity), and the colours (indicated in the key) represent eccentricities. 
All of our 25 orbits produce $m_{*,\,{\rm new}}\sim(2-6)\times10^{8}M_{\odot}$, with efficiencies ranging from $\sim$1.5$-$3.

It is beyond the scope of this work to identify exactly how star formation triggering depends on $\epsilon$ and $b$ \citep[see, e.g.,][]{dimatteo07}. In broad terms, the amount of star formation is governed by two factors: the strength of the interaction and its duration. In particular, highly eccentric orbits with large impact parameters last longer, leading to larger values of $m_{*,\,{\rm new}}$ (top panel). Correcting for the duration of the orbit (by dividing by $m^{\rm iso}_{*,\,{\rm new}}$) shows that low eccentricities and intermediate impact parameters lead to the highest star-formation efficiencies (bottom panel).

\subsubsection{Nuclear versus Off-Nuclear Efficiency}\label{subsubsecLocalEff}

We split our galaxies into two parts: the {\bf nuclear region} ($R < 1$ kpc) and the {\bf off-nuclear region} ($1 < R < 10$ kpc). Figure~\ref{figLocalEIP} is analogous to Figure~\ref{figGlobalEIP}, but  constrained to the nuclear (left) and off-nuclear (right panels) regions. 

The {\bf fraction of stellar mass} per region, defined as
\begin{equation}
{\rm f}_{\rm SFR} = \frac{\int_{\rm region} 4\pi R^2\,\rho_{*,\,{\rm new}}(R) \,{\rm d} R}{\int^{10\,{\rm kpc}}_{0} 4\pi R^2\,\rho_{*,\,{\rm new}}(R) \,{\rm d} R},
\label{eqnfrac}
\end{equation}
is shown in the middle panels.  $\rho_{*,\,{\rm new}}(R)$ is the mass-density radial profile of new stars created throughout the interaction. Our 25 mergers produce $m_{*,\,{\rm new}}\sim(1.5-5)\times10^{8}M_{\odot}$ in the central kpc (upper left), and $\sim(0.5-1)\times10^{8}M_{\odot}$ in the outskirts (upper right). Across all orbits (all values of $\epsilon$ and $b$), the mass in new stars is $\sim$87\% in the nucleus (middle left), and $\sim$13\% elsewhere (middle right). These portions are nearly independent of $\epsilon$ and $b$ (middle panels).

In stark contrast to the global case ({Figure~\ref{figGlobalEIP}), star formation efficiency is not above unity across the entire galaxy. Instead, it ranges between $\sim$4-10 in the nucleus and $\sim$0.2$-$0.5 in the outskirts. In other words, interactions enhance star formation in the centre, and suppress it at large radii. In particular, the suppression found in our case study (Figure~\ref{figLocalFiducial}) is generic across all of our 25 merger simulations.

\begin{figure}
 \centering
 \includegraphics[width=\columnwidth]{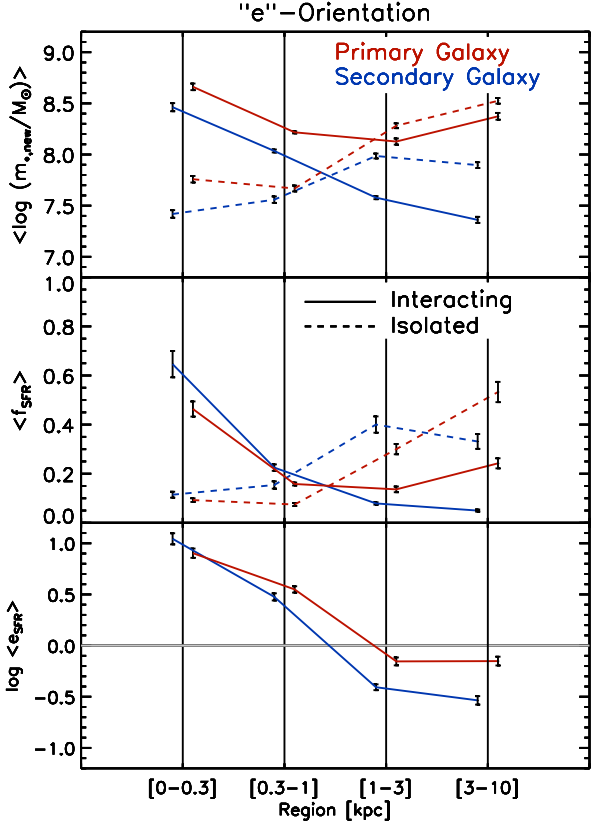}
 \caption{Mean mass in new stars, $\langle m_{*,\,{\rm new}}\rangle$ ({\it top}), mean SFR fraction $\langle {\rm f}_{\rm SFR} \rangle$, equation~\ref{eqnmeanfrac}, {\it middle}), and mean SFR efficiency, $\langle {\rm e}_{\rm SFR} \rangle$ (equation~\ref{eqnmeaneff}, {\it bottom})-- across 25 merger simulations (interacting phase only), with varying $\epsilon$ (eccentricity) and $b$ (impact parameter). We only include orbits in the ``e" orientation. Blue (red) lines correspond to the secondary (primary) galaxy. Grey horizontal line refers to mean SFR efficiency of unity. Solid vertical lines indicate the following regions (concentric spherical shells): $[0-0.3]$, $[0.3-1]$, $[1-3]$ and $[3-10]$ kpc. Symbols are displaced for clarity. Both galaxies have centrally-concentrated star formation, enhanced efficiencies in the central regions (first two bins), and suppressed efficiencies (below unity) everywhere else. Trends are more evident for the secondary galaxy.} 
 \label{figPrimary}
\end{figure}

\subsection{Comparing the Two Galaxies}\label{subsecPrimary}

Figures~\ref{figGlobalEIP} and \ref{figLocalEIP} demonstrate that quantities like $m_{*,\,{\rm new}}$, ${\rm f}_{\rm SFR}$, and ${\rm e}_{\rm SFR}$ are weakly dependent on $\epsilon$ and $b$.  Hereafter we only report averages over our 25 simulations (denoted in brackets). Error bars represent standard deviation of the mean. We now include the primary galaxy in our analysis.

We define the {\bf mean star-formation efficiency} as
\begin{equation}
\langle {\rm e}_{\rm SFR} \rangle =\frac{\langle m_{*,\,{\rm new}}\rangle}{\langle m^{\rm iso}_{*,\,{\rm new}}\rangle},
\label{eqnmeaneff}
\end{equation}
and the {\bf mean star-formation fraction} as
\begin{equation}
\langle {\rm f}_{\rm SFR} \rangle= \frac{\langle \int_{\rm region} 4\pi R^2\,\rho_{*,\,{\rm new}}(R) \,{\rm d} R \rangle}{\langle \int^{10\,{\rm kpc}}_{0} 4\pi R^2\,\rho_{*,\,{\rm new}}(R) \,{\rm d} R \rangle}.
\label{eqnmeanfrac}
\end{equation}
(Strictly speaking, $\langle {\rm f}_{\rm SFR} \rangle$ and $\langle {\rm e}_{\rm SFR} \rangle$ are not proper averages, but ratios of averaged quantities.)  We concentrate on four spherical shells spanning the following radii: $[0-0.3]$, $[0.3-1]$, $[1-3]$ and $[3-10]$ kpc (as in Figure~\ref{figLocalFiducial}).

Figure~\ref{figPrimary} shows $\langle m_{*,\,{\rm new}} \rangle$ (upper panel), $\langle {\rm f}_{\rm SFR} \rangle$ (middle panel) and $\langle {\rm e}_{\rm SFR} \rangle$ (bottom panel). Blue (red) lines represent the secondary (primary) galaxy -- solid (dashed) lines correspond to interacting (isolated) galaxies. Vertical lines mark the four spatial regions of interest, and values are offset for clarity.  
For both galaxies, the presence of a companion makes star formation more centrally concentrated. The bottom panel shows that $\langle {\rm e}_{\rm SFR} \rangle$ is enhanced in the inner regions, and {\it suppressed} in the outer shells. This is particularly evident for the secondary galaxy (by a factor of $\sim$0.3 in the last shell), and less obvious for the primary galaxy (by a factor of $\sim$0.7 in that same shell), suggesting that the smaller galaxy is more susceptible to interaction-driven effects.

\subsection{Disc Spin Orientation}\label{subsecOrientation}

\begin{figure}
 \centering
 \includegraphics[width=\columnwidth]{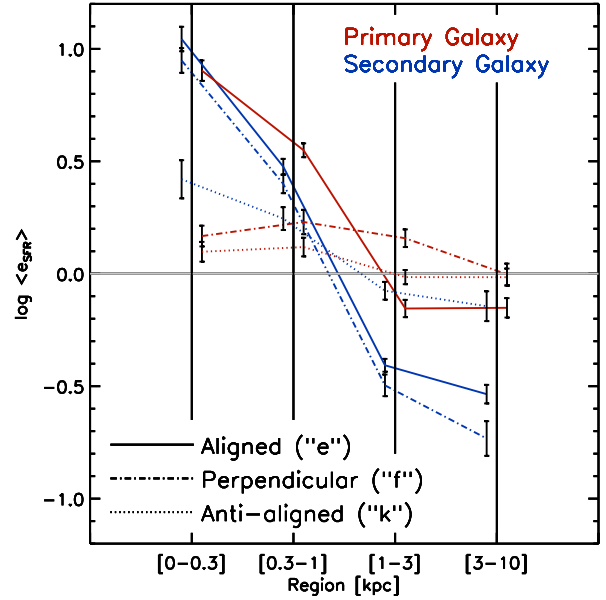}
 \caption{Mean star formation efficiency, $\langle {\rm e}_{\rm SFR}\rangle$ (equation~\ref{eqnmeaneff}), across 75 merger simulations (interaction stages only). Blue (red) lines represent the secondary (primary) galaxy. Grey horizontal line refers to mean SFR efficiency of unity. Vertical lines indicate the following regions (concentric spherical shells): $[0-0.3]$, $[0.3-1]$, $[1-3]$ and $[3-10]$ kpc. Symbols are displaced for clarity. Three orientations are considered (25 orbits each): ``e" (aligned), ``f" (perpendicular), ``k" (anti-aligned). See Figure~\ref{figDiskOrientation} and Table~\ref{tableAngles} for definitions. Both the ``e" and ``f" produce centrally-concentrated star formation in the secondary galaxy. Only the ``e" orientation has this effect on the primary. For anti-aligned galaxies (``k" orientation), the effect on either galaxy is minimal.} 
 \label{figOrientation}
\end{figure}

Figure~\ref{figOrientation} shows the mean star formation efficiency (equation~\ref{eqnmeaneff}) for the ``e" (solid lines, aligned discs), ``f" (dot-dashed lines) and ``k" (dotted lines) orientations. These choices represent (nearly) aligned, perpendicular and anti-aligned discs (Figure~\ref{figDiskOrientation} and Table~\ref{tableAngles}). The solid lines here are identical to the bottom panel of Figure~\ref{figPrimary}. 

For the secondary galaxy (blue), the ``e" and ``f" orientations behave similarly: star formation efficiency is enhanced in the centre, and suppressed in the outskirts. In contrast, the ``k" orientation exhibits substantially weaker enhancement (suppression) in the nucleus (outskirts). 
For the primary galaxy (red), the ``e" orientation is the only one showing strong enhanced (suppressed) star formation efficiency in the nucleus (outskirts). The other two orientations, are weakly enhanced in the central regions. The ``f" orientation is also weakly enhanced in the outskirts, whilst the ``k" orientation is consistent with unity in that region. 

The ``k" orientation (both galaxies) and ``f" orientation (primary galaxy) exhibit the weakest deviations from the isolated setting. After checking individual cases, these deviations are caused primarily by orbits with $b=1,2$, which allow  disc interprenetration at first pericentric passage.


\section{Discussion \& Future Work}\label{secDiscussion}

\subsection{Emerging Picture}\label{secPicture}

Standard wisdom suggests that galaxy encounters trigger bursts of star formation \citep{hernquist89,barnes91,barnes96}. However, the spatial extent of such episodes, and whether or not these events occur in both galaxies, remain poorly understood. To address these issues, we employ a suite of 75 merger simulations. We encapsulate the level of star formation in each orbital configuration in terms of the star-forming efficiency, defined as the ratio of the mass created during the interaction and the equivalent amount created in isolation ($\langle {\rm e}_{\rm SFR}\rangle$, equation~\ref{eqnmeaneff}). We find that, for the entire galaxy (Figure~\ref{figGlobalEIP}) and for individual shells (Figure~\ref{figLocalEIP}), this quantity is weakly dependent on $\epsilon$ and $b$ -- at least for the range of values explored here \citep[selected to be consistent with cosmological simulations,][]{khochfar06}.

The intensity and spatial extent of star formation in interacting galaxies is strongly driven by the alignment of the two disc spins (Figure~\ref{figOrientation}). With the exception of a few anomalous cases where the two galaxies interpenetrate one another, interactions with anti-aligned spins (``k"-orientation) have a minimal effect on the participating galaxies. The alignment of the two spins, however, permits centrally-concentrated star formation. The secondary galaxy experiences nuclear starbursts with either perpendicular (``f"-orientation) or aligned (``e"-orientation) spins. The primary galaxy, on the other hand, requires the two spins to be aligned for nuclear star formation to occur. 

Our results suggest that relative spin disc orientation is the dominant factor behind centrally-concentrated star formation in interacting galaxies, particularly for the smaller galaxy. Interestingly, we find that whenever enhanced star formation in the nucleus is triggered, this is always accompanied by suppression of activity at large galacto-centric radii. The detailed physics behind this process is explored in forth-coming work (Moreno et al., in prep).

\subsection{Connection to Other Work}\label{secOther}

Little systematic work has been done to quantify the spatial distribution of star formation in mergers (i.e., by exploring a broad range of orbital parameters). \cite{dimatteo07} is an exception. Unfortunately, these authors only report maps at a few time snapshots, with limited quantitative information on how intense this process is at each region. 

Interestingly, a large number of observational and theoretical works emphasize the importance of widespread star formation -- both in tidal tails and bridges. These include the {\it Antennae Galaxies NGC 4038/39} \citep{mirabel98,wang04,renaud08,brandl09,karl10,teyssier10,whitmore10,zhang10,karl11,karl13,privon13,renaud14,renaud15}; the {\it Mice 4676A/B} \citep{sotnikova98,read03,barnes04,privon13,wild14}; {\it Arp 299 IC 694/NGC 3690} \citep{alonsoherrero00,alonsoherrero09,sliwa12}; {\it NGC 1512/1510} \citep{koribalski09,ducci14}; {\it NGC 2207/ IC 2163} \citep{elmegreen00,struck05,elmegreen06,kaufman12,mineo13}; and the {\it Condor NGC 6872} \citep{eufrasio14}. Whilst compelling, it is not clear if the lessons learnt from these specific cases can be generalised.

One could argue that tail/bridge star formation is ubiquitous. In our simulations, this only occurs for orbits where the two discs interpenetrate, and only for a few Myr. Our analysis suggests that this mode of SFR localization is subdominant (Figure~\ref{figOrientation}). This is consistent with observations. \cite{schmidt13} embarked on a systematic study of 60 merging systems (\textsc{3d-hst}). They show that only $\sim$3\% of their mergers exhibit star formation in the regions ``in between" -- showing that, whilst this phenomenon does exist, it is rather uncommon. For this reason, we do not attempt to refine our underlying modelling, nor do we try to compare against any of the above specialised studies.

It is worth noting that \cite{schmidt13} also compare to simulations, which happen to be very similar to ours \citep{cox06,cox08}. However, they claim that their simulations do not agree with observations. They report more instances with both galaxies exhibiting elevated levels of star formation simultaneously in the simulations than in their observations ($\sim$59\% versus $\sim$32\%). They argue that this is due to the fact that both galaxies in their simulations have identical initial gas fractions; which might not be the case for real interacting galaxies. This interpretation is certainly viable. Another explanation is that their simulations and observations could be covering a non-representative handful of time snapshots and projections (observational viewpoints) -- which might be incapable of fully capturing the interaction-induced star-forming process in these systems (60 observed and 296 simulated maps). In our simulations we see situations where both galaxies begin with the same gas fraction, but exhibit different levels of star formation at a given point in time. For instance, in our case study,  the primary galaxy is still forming stars during the late stages of the interacting phase, whilst the companion is not (upper panel of Figure~\ref{figGlobalFiducial}). 

\subsection{Future Directions}\label{secFuture}

There are several directions worth pursuing. Figure~\ref{figMaps} shows the complex morphology of star forming gas in interacting galaxies. In particular, the ring-like structure appearing after first  passage is intriguing. If robust, this imprint could provide a promising method for identifying galaxies that have experienced a close encounter in the past -- e.g., compare columns (c) and (d) in Figure~\ref{figMaps} to column (a). This could prove to be a powerful alternative to selecting interactions via the presence of tidal tails \citep{kartaltepe12,darg10a,casteels13,hung13,hung14}, which dissipate faster than this ring-like feature (column b) -- provided that observations are capable of measuring these detailed features (via H$\alpha$ and [OIII] emission-line maps). We acknowledge that at this point, it is not clear if the properties of this feature (or its existence and observability) depend strongly on numerical/modeling effects. This warrants a more rigorous study. The conclusions of this paper are not affected by this ring, which only accounts for $\sim$4\% of the produced stars in our simulations. 

We also plan to construct a `mock survey' -- along the lines of \cite{schmidt13}, but with more snapshots, viewpoints, and merger configurations. This framework will replace three-dimensional spherical shells with cylindrical annuli, to compare better with observations. Azimuthal dependences within these annuli will capture tidal tails and other features. This mock survey will be better suited for capturing the localization of SFR on a snapshot-by-snapshot basis -- thereby overcoming some of the limitations inherent to the time-integrated metrics (e.g., $e_{\rm SFR}$, equation~\ref{eqneff}) employed in this paper. 

Our results allude to the importance of exploring other orientations and mass ratios. We find that the smaller galaxy has a more evident (nuclear enhanced, off-nuclear suppressed) response to the encounter.  With this, we speculate that this trend will continue for other mass ratios (beyond our adopted choice, $\sim$2.5:1). That is, for more discrepant ratios, nuclear triggering in smaller galaxy might require weaker alignments, and stronger alignments for the larger galaxy. Clearly, this requires exploring a more refined range of orientations, and extending our mass-ratio regime.

Ideally, it is also desirable to explore other mass, gas fraction, and bulge-to-disc ratio combinations. We must employ extra care because massive galaxies tend to be more bulge-dominated \citep[e.g.,][]{kauffmann03,bluck14}, and have lower gas fractions \citep[e.g.,][]{stewart09,catinella12}. Furthermore, bulge-dominated systems tend to inhabit denser environments \citep{dressler80,butcher84}, rendering our ``isolated galaxy merger" approximation less valid \citep{martig08,moreno12,moreno13}. 


\section{Conclusions}\label{secConclusions}

The aim of this work is to investigate the spatial extent of star formation in interacting galaxies. To address this, we employ a suite of 75 idealised SPH merger simulations (with stellar masses $M_{*}=14 \times 10^{10} M_{\odot}$ and $5.7 \times 10^{9} M_{\odot}$, bulge-to-disc mass equal to 0.31, and gas fraction equal to 0.25). We only consider three relative spin orientations, labelled ``e", ``f" and ``k".  These choices are meant to represent two nearly aligned discs (``e"-orientation), two perpendicular discs (``f"-orientation), and two anti-aligned discs (``k''-orientation). For each fixed spin disc orientation, we explore five eccentricities ($\epsilon=\{0.85, 0.90, 0.95, 1.0, 1.05 \}$) and five impact parameters ($b=\{2, 4, 8, 12, 16\}$ kpc). 

In this work, we quantify the spatial distribution of star formation in terms of spherical shells of radii $[0 - 0.3]$, $[0.3 - 1]$, $[1 - 3]$ and $[3 - 10]$ kpc. We focus exclusively on the interacting stage, between first and second pericentric passage. We employ star-formation efficiency (equation~\ref{eqneff}) -- defined as the integrated star formation during the interacting phase, divided by its analogue in isolation -- to encapsulate interaction-induced effects for each orbit. We warn that our results only apply to the period when the two merging galaxies can still be identified as separate entities.\\

Our main results are:
\begin{itemize}

\item Interactions generally produced {\it enhanced} star formation in the centre, and {\it suppressed} activity in the outskirts. This effect is weakly dependent on the values of $\epsilon$ and $b$ probed here (Figures~\ref{figGlobalEIP}).

\item These trends are more pronounced in interactions with strongly aligned disc spin {\it orientations} (Figure~\ref{figOrientation}), particularly for the {\it secondary} (smaller) galaxy (Figure~\ref{figPrimary}).

\end{itemize}

It is our hope that the numerical investigations presented in this paper motivate detailed observational studies with samples of galaxies drawn from ongoing and future deep-field and integral-field-spectroscopic surveys.


\section*{acknowledgments}

The computations in this paper were run on the Odyssey cluster supported by the FAS Division of Science, Research Computing Group at Harvard University. JM acknowledges the Canadian Institute for Theoretical Astrophysics for partial funding, and Phil Hopkins for being a wonderful host towards the end of this project. JM, SE, DP and AB are funded by the Natural Science and Engineering Research. GB thanks MITACS for making her stay in Victoria possible. The authors thank the referee, Fr\'{e}d\'{e}ric Bournaud, for a timely review that greatly improved this paper -- as well as Phil Hopkins and Florent Renaud for useful discussions on an earlier draft. JM thanks the organisers and participants of {\it 3D2014: Gas and stars in galaxies: A multi-wavelength 3D perspective} (Munich, March 2014) and {\it Galaxies in 3D across the universe} (Vienna, July 2014) -- for instigating incredibly engaging discussions on this subject. 




\bibliographystyle{mn2e}

\begin{thebibliography}{}

\bibitem[Alonso et al.(2004)]{alonso04} Alonso M.~S., Tissera P.~B., Coldwell G., Lambas D.~G.,\ 2004, MNRAS, 352, 1081

\bibitem[Alonso et al.(2006)]{alonso06} Alonso M.~S., Lambas D.~G., Tissera P., Coldwell G.,\ 2006, MNRAS, 367, 1029

\bibitem[Alonso et al.(2012)]{alonso12} Alonso S., Mesa V., Padilla N., Lambas D.~G.,\ 2012, A\&A, 539, A46 

\bibitem[Alonso-Herrero et al.(2000)]{alonsoherrero00} Alonso-Herrero A., Rieke G.~H., Rieke M.~J., Scoville N.~Z.,\ 2000, ApJ, 532, 845 

\bibitem[Alonso-Herrero et al.(2009)]{alonsoherrero09} Alonso-Herrero A., Rieke G.~H., Colina L., et al.,\ 2009, ApJ, 697, 660 

\bibitem[Barnes \& Hernquist(1991)]{barnes91} Barnes J.~E., Hernquist L.~E.,\ 1991, ApJL, 370, L65

\bibitem[Barnes \& Hernquist(1996)]{barnes96} Barnes J.~E., Hernquist L.,\ 1996, ApJ, 471, 115

\bibitem[Barnes(2004)]{barnes04} Barnes J.~E.,\ 2004, MNRAS, 350, 798 

\bibitem[Barton et al.(2000)]{barton00} Barton E.~J., Geller M.~J., Kenyon S.~J.,\ 2000, ApJ, 530, 660 

\bibitem[Barton Gillespie et al.(2003)]{barton03} Barton Gillespie E., Geller M.~J., Kenyon S.~J.\ 2003, ApJ, 582, 668 

\bibitem[Bellocchi et al.(2013)]{bellocchi13} Bellocchi E., Arribas S., Colina L., Miralles-Caballero D.,\ 2013, A\&A, 557, A59 

\bibitem[Bluck et al.(2014)]{bluck14} Bluck A.~F.~L., Mendel J.~T., Ellison, S.~L., et al.,\ 2014, MNRAS, 441, 599

\bibitem[Brandl et al.(2009)]{brandl09} Brandl B.~R., Snijders L., den Brok M., et al.,\ 2009, ApJ, 699, 1982

\bibitem[Bundy et al.(2015)]{bundy15} Bundy K., Bershady M.~A., Law D.~R., et al.,\ 2015, ApJ, 798, 7

\bibitem[Butcher \& Oemler(1984)]{butcher84} Butcher H., Oemler A., Jr.,\ 1984, ApJ, 285, 426

\bibitem[Catinella et al.(2012)]{catinella12} Catinella B., Schiminovich D., Kauffmann G., et al.,\ 2012, A\&A, 544, AA65 

\bibitem[Casteels et al.(2013)]{casteels13} Casteels K.~R.~V., Bamford S.~P., Skibba R.~A., et al.,\ 2013, MNRAS, 429, 1051

\bibitem[Cox et al.(2006)]{cox06} Cox T.~J., Jonsson P., Primack J.~R., Somerville R.~S.,\ 2006, MNRAS, 373, 1013 

\bibitem[Cox et al.(2008)]{cox08} Cox T.~J., Jonsson P., Somerville R.~S., Primack J.~R., Dekel A.,\ 2008, MNRAS, 384, 386 

\bibitem[Croom et al.(2012)]{croom12} Croom S.~M., Lawrence J.~S., Bland-Hawthorn J., et al.,\ 2012, MNRAS, 421, 872 

\bibitem[Darg et al.(2010a)]{darg10a} Darg D.~W., Kaviraj S., Lintott C.~J., et al.,\ 2010, MNRAS, 401, 1043 

\bibitem[Darg et al.(2010b)]{darg10b} Darg D.~W., Kaviraj S., Lintott C.~J., et al.,\ 2010, MNRAS, 401, 1552 

\bibitem[De Propris et al.(2014)]{depropris14} De Propris R., Baldry I.~K., Bland-Hawthorn J., et al.,\ 2014, MNRAS, 444, 2200

\bibitem[Di Matteo et al.(2005)]{dimatteo05} Di Matteo T., Springel V., Hernquist L.,\ 2005, Nat, 433, 604 

\bibitem[Di Matteo et al.(2007)]{dimatteo07} Di Matteo P., Combes F., Melchior A.-L., Semelin B.,\ 2007, AAP, 468, 61

\bibitem[Di Matteo et al.(2008)]{dimatteo08} Di Matteo P., Bournaud F., Martig M., et al.,\ 2008, A\&A, 492, 31 

\bibitem[D'Onghia et al.(2010)]{donghia10} D'Onghia E., Vogelsberger M., Faucher-Giguere C.-A., Hernquist L.,\ 2010, ApJ, 725, 353 

\bibitem[Dressler(1980)]{dressler80} Dressler A.,\ 1980, ApJ, 236, 35

\bibitem[Ducci et al.(2014)]{ducci14} Ducci L., Kavanagh P.~J., Sasaki M., Koribalski B.~S.,\ 2014, A\&A, 566, A115

\bibitem[Ellison et al.(2008)]{ellison08} Ellison S.~L., Patton D.~R., Simard L., McConnachie A.~W.,\ 2008, AJ, 135, 1877

\bibitem[Ellison et al.(2010)]{ellison10} Ellison S.~L., Patton D.~R., Simard, L., McConnachie A.~W., Baldry I.~K., Mendel J.~T.,\ 2010, MNRAS, 407, 151

\bibitem[Ellison et al.(2011)]{ellison11} Ellison S.~L., Patton D.~R., Mendel J.~T., Scudder J.~M.,\ 2011, MNRAS, 418, 2043

\bibitem[Ellison et al.(2013)]{ellison13} Ellison S.~L., Mendel J.~T., Patton D.~R., Scudder J.~M.,\ 2013, MNRAS, 435, 3627

\bibitem[Elmegreen et al.(2000)]{elmegreen00} Elmegreen B.~G., Kaufman M., Struck C., et al.,\ 2000, AJ, 120, 630 

\bibitem[Elmegreen et al.(2006)]{elmegreen06} Elmegreen D.~M., Elmegreen B.~G., Kaufman M., et al.,\ 2006, ApJ, 642, 158

\bibitem[Eufrasio et al.(2014)]{eufrasio14} Eufrasio R.~T., Dwek E., Arendt R.~G., et al.,\ 2014, ApJ, 795, 89

\bibitem[Freedman Woods et al.(2010)]{freedman10} Freedman Woods D., Geller M.~J., Kurtz M.~J., et al.,\ 2010, AJ, 139, 1857 

\bibitem[Hayward et al.(2014)]{hayward14} Hayward C.~C., Torrey P., Springel V., Hernquist L., Vogelsberger M.,\ 2014, MNRAS, 442, 1992 

\bibitem[Hernquist(1990)]{hernquist90} Hernquist L.,\ 1990, ApJ, 356, 359 

\bibitem[Hernquist(1989)]{hernquist89} Hernquist L.,\ 1989, Nat, 340, 687

\bibitem[Hopkins et al.(2013)]{hopkins13} Hopkins P.~F., Cox T.~J., Hernquist L., et al.,\ 2013, MNRAS, 430, 1901

\bibitem[Hung et al.(2013)]{hung13} Hung C.-L., Sanders D.~B., Casey C.~M., et al.,\ 2013, ApJ, 778, 129 

\bibitem[Hung et al.(2014)]{hung14} Hung C.-L., Sanders D.~B., Casey C.~M., et al.,\ 2014, ApJ, 791, 63

\bibitem[Hwang et al.(2011)]{hwang11} Hwang H.~S., Elbaz D., Dickinson M., et al.,\ 2011, A\&A, 535, A60 

\bibitem[Iono et al.(2004)]{iono04} Iono D., Yun M.~S., Mihos J.~C.,\ 2004, ApJ, 616, 199 

\bibitem[Kampczyk et al.(2013)]{kampczyk13} Kampczyk P., Lilly S.~J., de Ravel L., et al.,\ 2013, ApJ, 762, 43

\bibitem[Karl et al.(2010)]{karl10} Karl S.~J., Naab T., Johansson P.~H., et al.,\ 2010, ApJL, 715, L88 

\bibitem[Karl et al.(2011)]{karl11} Karl S.~J., Fall S.~M., Naab T.,\ 2011, ApJ, 734, 11 

\bibitem[Karl et al.(2013)]{karl13} Karl S.~J., Lunttila T., Naab T., et al.,\ 2013, MNRAS, 434, 696 

\bibitem[Kartaltepe et al.(2012)]{kartaltepe12} Kartaltepe J.~S., Dickinson M., Alexander D.~M., et al.,\ 2012, ApJ, 757, 23 

\bibitem[Katz et al.(1996)]{katz96} Katz N., Weinberg D.~H., Hernquist L.,\ 1996, ApJS, 105, 19 

\bibitem[Kauffmann et al.(2003)]{kauffmann03} Kauffmann G., Heckman T.~M., White S.~D.~M., et al.\ 2003, MNRAS, 341, 54 

\bibitem[Kaufman et al.(2012)]{kaufman12} Kaufman M., Grupe D., Elmegreen B.~G., et al.,\ 2012, AJ, 144, 156 

\bibitem[Kennicutt(1998)]{kennicutt98} Kennicutt R.~C., Jr.,\ 1998, ApJ, 498, 541 

\bibitem[Kewley et al.(2010)]{kewley10} Kewley L.~J., Rupke D., Zahid H.~J., Geller M.~J., Barton E.~J.,\ 2010, ApJL, 721, L48

\bibitem[Koribalski \& L{\'o}pez-S{\'a}nchez(2009)]{koribalski09} Koribalski B.~S., L{\'o}pez-S{\'a}nchez {\'A}.~R.,\ 2009, MNRAS, 400, 1749

\bibitem[Knapen \& James(2009)]{knapen09} Knapen J.~H., James P.~A.,\ 2009, ApJ, 698, 1437 

\bibitem[Knapen et al.(2014)]{knapen14} Knapen J.~H., Erroz-Ferrer S., Roa J., et al.,\ 2014, A\&A, 569, AA91

\bibitem[Khochfar \& Burkert(2006)]{khochfar06} Khochfar S., Burkert A.,\ 2006, A\&A, 445, 403 

\bibitem[Lambas et al.(2003)]{lambas03} Lambas D.~G., Tissera P.~B., Alonso M.~S., Coldwell G.,\ 2003, MNRAS, 346, 1189 

\bibitem[Lawrence et al.(2012)]{lawrence12} Lawrence J., Bland-Hawthorn J., Bryant J., et al.,\ 2012, PROCSPIE, 8446,  

\bibitem[Lin et al.(2007)]{lin07} Lin L., Koo D.~C., Weiner B.~J., et al.,\, 2007, ApJL 660, L51 

\bibitem[Martig \& Bournaud(2008)]{martig08} Martig M., Bournaud F.,\ 2008, MNRAS, 385, L38 

\bibitem[Mihos \& Hernquist(1996)]{mihos96} Mihos J.~C., Hernquist L.,\ 1996, ApJ, 464, 641 

\bibitem[Mineo et al.(2013)]{mineo13} Mineo S., Rappaport S., Steinhorn B., et al.,\ 2013, ApJ, 771, 133

\bibitem[Mirabel et al.(1998)]{mirabel98} Mirabel I.~F., Vigroux L., Charmandaris V., et al.,\ 1998, A\&A, 333, L1

\bibitem[Mo et al.(1998)]{mo98} Mo H.~J., Mao S., White S.~D.~M.,\ 1998, MNRAS, 295, 319

\bibitem[Moreno(2012)]{moreno12} Moreno J.,\ 2012, MNRAS, 419, 
411 

\bibitem[Moreno et al.(2013)]{moreno13} Moreno J., Bluck A.~F.~L., Ellison S.~L., et al.,\ 2013, MNRAS, 436, 1765

\bibitem[Patton et al.(2011)]{patton11} Patton D.~R., Ellison S.~L., Simard L., McConnachie A.~W., Mendel J.~T.,\ 2011, MNRAS, 412, 591 

\bibitem[Patton et al.(2013)]{patton13} Patton D.~R., Torrey P., Ellison S.~L., Mendel J.~T., Scudder J.~M.,\ 2013, MNRAS, 433, L59 

\bibitem[Powell et al.(2013)]{powell13} Powell L.~C., Bournaud F., Chapon D., Teyssier R.,\ 2013, MNRAS, 434, 1028 

\bibitem[Privon et al.(2013)]{privon13} Privon G.~C., Barnes J.~E., Evans A.~S., et al.,\ 2013, ApJ, 771, 120 

\bibitem[Read(2003)]{read03} Read A.~M.,\ 2003, MNRAS, 342, 715 

\bibitem[Renaud et al.(2008)]{renaud08} Renaud F., Boily C.~M., Fleck J.-J., Naab T., Theis C.,\ 2008, MNRAS, 391, L98 

\bibitem[Renaud et al.(2014)]{renaud14} Renaud F., Bournaud F., Kraljic K., Duc P.-A.,\ 2014, MNRAS, 442, L33

\bibitem[Renaud et al.,(2015)]{renaud15} Renaud F., Bournaud F., \& Duc, P.-A.,\ 2015, MNRAS, 446, 2038

\bibitem[Rich et al.(2012)]{rich12} Rich J.~A., Torrey P., Kewley L.~J., Dopita M.~A., Rupke D.~S.~N.,\ 2012, ApJ, 753, 5 

\bibitem[Robertson et al.(2006)]{robertson06} Robertson B., Bullock J.~S., Cox T.~J., Di Matteo T., Hernquist L., Springel V., Yoshida N.,\ 2006, ApJ, 645, 986 

\bibitem[Rosa et al.(2014)]{rosa14} Rosa D.~A., Dors O.~L., Krabbe A.~C., et al.,\ 2014, MNRAS, 444, 2005 

\bibitem[Rupke et al.(2010)]{rupke10} Rupke D.~S.~N., Kewley L.~J., Chien L.-H.,\ 2010, ApJ, 723, 1255

\bibitem[S{\'a}nchez et al.(2014)]{sanchez14} S{\'a}nchez S.~F., Rosales-Ortega F.~F., Iglesias-P{\'a}ramo J., et al.,\ 2014, A\&A, 563, A49 

\bibitem[Sanders \& Mirabel(1996)]{sanders96} Sanders D.~B., Mirabel I.~F.,\ 1996, ARAA, 34, 749

\bibitem[Schmidt et al.(2013)]{schmidt13} Schmidt K.~B., Rix H.-W., da Cunha E., et al.,\ 2013, MNRAS, 432, 285

\bibitem[Scott \& Kaviraj(2014)]{scott14} Scott C., Kaviraj S.,\ 2014, MNRAS, 437, 2137 

\bibitem[Scudder et al.(2012)]{scudder12} Scudder J.~M., Ellison S.~L., Torrey P., Patton D.~R., Mendel J.~T.,\ 2012, MNRAS, 426, 549

\bibitem[Sliwa et al.(2012)]{sliwa12} Sliwa K., Wilson C.~D., Petitpas G.~R., et al.,\ 2012, ApJ, 753, 46

\bibitem[Sheth et al.(2010)]{sheth10} Sheth K., Regan M., Hinz J.~L., et al.,\ 2010, PASP, 122, 1397 

\bibitem[Sotnikova \& Reshetnikov(1998)]{sotnikova98} Sotnikova N.~Y., Reshetnikov V.~P.,\ 1998, Astronomy Letters, 24, 73 

\bibitem[Springel(2000)]{springel00} Springel V.\ 2000, MNRAS, 312, 859

\bibitem[Springel \& Hernquist(2002)]{springel02} Springel V., Hernquist L.,\ 2002, MNRAS, 333, 649

\bibitem[Springel \& Hernquist(2003)]{springel03} Springel V., Hernquist L.,\ 2003, MNRAS, 339, 289 

\bibitem[Springel(2005)]{springel05gadget} Springel V.,\ 2005, MNRAS, 364, 1105 

\bibitem[Springel et al.(2005)]{springel05} Springel V., Di Matteo T., Hernquist L.,\ 2005, MNRAS, 361, 776 

\bibitem[Springel \& Hernquist(2005)]{springel05b} Springel V., Hernquist L.,\ 2005, ApJL, 622, L9 

\bibitem[Springel et al.(2005b)]{springel05ms} Springel V., et al.,\  2005, Nat, 435, 629

\bibitem[Stewart et al.(2009)]{stewart09} Stewart K.~R., Bullock J.~S., Wechsler R.~H., Maller A.~H.,\ 2009, ApJ, 702, 307 

\bibitem[Struck et al.(2005)]{struck05} Struck C., Kaufman M., Brinks E., et al.,\ 2005, MNRAS, 364, 69 

\bibitem[Teyssier et al.(2010)]{teyssier10} Teyssier R., Chapon D., Bournaud F.\, 2010, ApJL, 720, L149 

\bibitem[Torrey et al.(2012)]{torrey12} Torrey P., Cox T.~J., Kewley L., Hernquist L.,\ 2012, ApJ, 746, 108 

\bibitem[Wang et al.(2004)]{wang04} Wang Z., Fazio G.~G., Ashby M.~L.~N., et al.,\ 2004, ApJS, 154, 193 

\bibitem[Whitmore et al.(2010)]{whitmore10} Whitmore B.~C., Chandar R., Schweizer F., et al.,\ 2010, AJ, 140, 75 

\bibitem[Wild et al.(2014)]{wild14} Wild V., Rosales-Ortega F., Falcon-Barroso J., et al.,\ 2014, A\&A, 567, AA132 

\bibitem[Wong et al.(2011)]{wong11} Wong K.~C., Blanton M.~R., Burles S.~M., et al.,\ 2011, ApJ, 728, 119 

\bibitem[Woods et al.(2006)]{woods06} Woods D.~F., Geller M.~J., \& Barton E.~J.,\, 2006, AJ, 132, 197

\bibitem[Zhang et al.(2010)]{zhang10} Zhang H.-X., Gao Y., Kong X.,\ 2010, MNRAS, 401, 1839

\end{thebibliography}

\label{lastpage}
\end{document}